\documentclass[12pt,tightenlines,eqsecnum,floats,showpacs,nofootinbib,amsmath,amssymb,aps,prd,superscriptaddress]{revtex4-1}

\usepackage{graphicx}
\usepackage{amsmath,amssymb}
\usepackage{hyperref}

\begin{document}

\title{Approximation methods in Loop Quantum Cosmology: From Gowdy
cosmologies to inhomogeneous models in Friedmann--Robertson--Walker geometries.}

\author{Mercedes Mart\'in-Benito}
\email{mmartinbenito@perimeterinstitute.ca}
\affiliation{Perimeter Institute for Theoretical Physics, 31 Caroline Str. N., N2L 2Y6 Waterloo, Canada}
\author{Daniel \surname{Mart\'in-de~Blas}}
\email{daniel.martin@iem.cfmac.csic.es}
\affiliation{Instituto de Estructura de la Materia, IEM-CSIC, Serrano 121, 28006 Madrid, Spain}
\author{Guillermo A. Mena Marug\'an} \email{mena@iem.cfmac.csic.es}
\affiliation{Instituto de Estructura de la Materia, IEM-CSIC,
Serrano 121, 28006 Madrid, Spain}

\begin{abstract}
We develop approximation methods in the hybrid quantization of the Gowdy model with linear polarization and a massless scalar field, for the case of three-torus spatial topology. The loop quantization of the homogeneous gravitational sector of the Gowdy model (according to the improved dynamics prescription) and the presence of inhomogeneities lead to a very complicated Hamiltonian constraint. Therefore, the extraction of physical results calls for the introduction of well justified approximations. We first show how to approximate the homogeneous part of the Hamiltonian constraint, corresponding to Bianchi I geometries, as if it described a Friedmann--Robertson--Walker (FRW) model corrected with anisotropies. This approximation is valid in the sector of high energies of the FRW geometry (concerning its contribution to the constraint) and for anisotropy profiles that are sufficiently smooth. In addition, for certain families of states related to regimes of physical interest, with negligible quantum effects of the anisotropies and small inhomogeneities, one can approximate the Hamiltonian constraint of the inhomogeneous system by that of an FRW geometry with a relatively simple matter content, and then obtain its solutions.
\end{abstract}


\maketitle

\section{Introduction}
\label{sec:IM}

One of the main challenges of modern physics is to construct a theory for Quantum Gravity, and in particular a theory for Quantum Cosmology. Indeed, General Relativity breaks down in the very initial
instants of the history of the universe, leading to a cosmological big bang singularity \cite{wald1}. In that regime, quantum geometry effects are expected to be important and should be taken into account.

Loop Quantum Cosmology (LQC) \cite{lqc1,lqc2,lqc3,lqc4} is a promising quantum approach for cosmological systems inspired by the ideas and methods of Loop Quantum Gravity  \cite{lqg1,lqg2,lqg3}. It has been applied mainly to---homogeneous and isotropic---Friedmann-Robertson-Walker (FRW) models with successful results \cite{aps,hom1,hom2,acs,hom3,hom4,hom4b,mmo,mop,su11,hom5}. Remarkably, it predicts a quantum bounce mechanism that eludes the initial big bang singularity \cite{aps}. To deal with more complicated systems, and to test the robustness of the results, LQC has also been applied to anisotropic models \cite{anis1,mbmmp,anis2,anis3,awe1,anis4}, as well as to inhomogeneous ones. The best studied inhomogeneous system is the vacuum $T^3$ Gowdy model with linear polarization. This is a midisuperspace with three-torus spatial topology that contains inhomogeneities (corresponding to gravitational waves) varying in a single direction \cite{gowdy}. The quantization of this model has been carried out adopting a hybrid approach, which combines techniques of LQC when representing the homogeneous sector of the model with a Fock quantization of the inhomogeneities \cite{hybrid1,hybrid2,hybrid3,hybrid4}.

More realistic analyses call for the introduction of matter. In this paper, we will consider the $T^3$ Gowdy model minimally coupled to a massless scalar field with the same symmetries as the spacetime metric. This is probably the simplest inhomogeneous model which contains FRW cosmologies as a subset of homogeneous solutions (which otherwise are not present in the model: in vacuo, the only homogeneous and isotropic solution is flat space-time). Actually, this Gowdy model can be regarded as an FRW background of three-torus spatial topology with anisotropies and (gravitational and matter) inhomogeneities propagating on it. Though this scenario is not yet completely realistic, because the inhomogeneities vary along a single spatial direction, it provides a further step towards the quantum analysis of physical inhomogeneities in cosmology.

Owing to the isometries of the Gowdy model, the system admits a family of solutions with local rotational symmetry (LRS), in which the two scale factors of the directions of symmetry coincide. In order to simplify our analysis, we will focus on this type of solutions. The hybrid quantization of this LRS-Gowdy model was already described in Ref.~\cite{hybrid5}. The aim of the present work is to introduce well justified approximations to the quantum constraint of that model, keeping under control the conditions for which the approximations are valid, and then pass to consider regimes of physical interest where one can construct approximate solutions. Owing to the LQC techniques employed to represent the degrees of freedom of the homogeneous geometry, the Hamiltonian constraint of the quantum model is a quite complicated operator. Therefore, finding exact solutions to it is extremely difficult, if not impossible. The constraint contains mainly two types of terms which pose serious obstructions for its resolution. First, it includes a difference operator that couples the homogeneous and isotropic part of the background geometry with the anisotropies (we will call this the anisotropy term). Second, there appears an interaction term for the inhomogeneities (which is also coupled to the homogeneous and isotropic background and to the anisotropies). The anisotropy term has an intricate action. Our strategy will be to approximate it by another operator that, while retaining the discreteness and the main properties of the loop quantization, has a simpler action on the anisotropies, so that one can carry out its spectral analysis. Our approximation will be valid for states in the sector of high energies of the homogeneous and isotropic part of the background geometry\footnote{By ``high energy'' we mean large (eigen)values of the operator which provides the considered contribution to the quantum (densitized) Hamiltonian constraint.}, and with a suitably smooth profile for the anisotropies. Moreover, once this approximation is made, we will be able to find sets of states for which not only the contribution of this anisotropy term is negligible, but also the interaction term in the inhomogeneities can be disregarded. In this way, we will get a well justified approximation of the Hamiltonian constraint by a solvable one. From the solutions of this latter constraint, we will be able to construct approximate solutions for the Gowdy model.

These approximate solutions retain the interaction between the homogeneous and isotropic FRW background and the free energy of the inhomogeneities. Therefore these solutions are suitable, e.g., to conduct investigations about the robustness of the quantum bounce scenario of LQC when inhomogeneities are included, or about modifications in the evolution of the matter inhomogeneities when quantum geometry effects are taken into account. We leave for future work the analysis of those questions.

The organization of the paper is as follows. In Sec. \ref{sec:HGM} we summarize the main results of the hybrid quantization of the $T^3$ Gowdy model with LRS and minimally coupled to a massless scalar field. In Sec. \ref{sec:App} we introduce the approximations that are needed to go from the quantum Hamiltonian constraint of the Gowdy model to a solvable one in which the anisotropy term and the interaction term are not present. This includes approximating the Hamiltonian constraint of the Bianchi I homogeneous model (when no inhomogeneities are considered in the system) by the constraint of a homogeneous and isotropic FRW model corrected with some anisotropy contributions. The section contains some technical considerations about our approximations for the anisotropy term and the class of states for which they are valid. Then, in Sec. \ref{sec:sol} we construct approximate solutions for the Gowdy model. In Sec. \ref{sec:C} we conclude and present the main results of this work.

\section{Hybrid quantization of the Gowdy model}
\label{sec:HGM}

We consider the linearly polarized $T^{3}$ Gowdy model with LRS and minimally coupled to a free massless scalar field $\Phi$ with the same symmetries as the metric. The hybrid quantization of this model has already been carried out by the authors in Ref. \cite{hybrid5}, by extending to the case with matter the hybrid quantization of the model in vacuo \cite{hybrid1,hybrid2,hybrid3,hybrid4}. In this section, we summarize the main steps of this quantization. We refer to Refs. \cite{hybrid1,hybrid2,hybrid3,hybrid4,hybrid5} for further details.

We call $\{\theta, \sigma, \delta\}$ the three orthogonal spatial coordinates for the Gowdy spacetimes, each of them defined on the circle. Because of the symmetries, matter and gravitational fields have spatial dependence only in e.g. $\theta$, so that they can be expanded in Fourier modes in this coordinate. Our starting point is the reduced classical system that results after a partial gauge fixing. The resulting reduced phase space consists in a homogeneous sector which coincides with the phase space of the Bianchi I model with LRS and minimally coupled to a homogeneous massless scalar $\phi$ (the zero mode of the matter field $\Phi$ in the introduced Fourier expansion), and in an inhomogeneous sector formed by the non-zero modes of the linearly polarized gravitational wave and of the matter field, together with their canonically conjugate momenta. This reduced model is subject to two global constraints: a momentum constraint generating rigid rotations in $\theta$ and a Hamiltonian constraint that generates time reparametrizations. Our hybrid approach consists in adopting a Fock quantization for both the gravitational and matter inhomogeneities, a standard Schr\"odinger representation for the homogeneous massless scalar field $\phi$, and a loop quantization within the improved dynamics scheme for the Bianchi I variables \cite{awe1}.

Let us first focus on the representation of the Bianchi I phase space.

\subsection{Loop quantization of the Bianchi I model}\label{sec:HGM1}

In the loop approach the phase space of this model is described by the three non-vanishing components (in a diagonal gauge) of the corresponding $SU(2)$ Ashtekar-Barbero connection,  $(c_\theta, c_{\sigma}, c_\delta)$, and by the three non-vanishing components of the densitized triad, $(p_\theta, p_{\sigma}, p_\delta)$. For the model with LRS under consideration we have to impose the restrictions $c_{\sigma}=c_\delta\equiv c_\perp$ and $p_{\sigma}=p_\delta\equiv p_\perp$. These variables have non-vanishing Poisson brackets  $\{c_\theta,p_\theta\}=8\pi G\gamma$ and $\{c_\perp,p_\perp\}=4\pi G\gamma$. Here $\gamma$ denotes the Immirzi parameter and $G$ the Newton constant. In these variables the Hamiltonian constraint of the LRS Bianchi I model coupled to $\phi$ reads
\begin{align}\label{chi}
C_{\text{BI}}= -\frac1{8\pi G\gamma^2}\left[2c_\theta p_\theta c_\perp p_\perp +( c_\perp p_\perp)^2\right] +\frac{p_\phi^2}{2},
\end{align}
where $p_\phi$ stands for the momentum canonically conjugate to $\phi$.

The improved dynamics scheme accounts for the existence of a minimum non-vanishing eigenvalue $\Delta$ of the area operator in LQG. This scheme involves first the introduction of the following variables  \cite{awe1}
\begin{align}\label{var}
\lambda_\theta=\frac{\sqrt{|p_\theta|}{\rm sign}(p_\theta)}{(4\pi G\hbar \gamma \sqrt{\Delta})^{1/3}},&\qquad b_\theta=\frac{\sqrt{\Delta|p_\theta|}}{p_\perp}c_\theta\equiv {\bar\mu}_\theta c_\theta,\nonumber\\
\qquad v=\frac{\sqrt{|p_\theta|}p_\perp{\rm sign}(p_\theta)}{2\pi G\hbar \gamma\sqrt{\Delta}}, &\qquad b=\sqrt{\frac{\Delta}{|p_\theta|}}{\rm sign}(p_\theta)c_\perp\equiv {\bar\mu}_\perp c_\perp.
\end{align}
We note that $v$ is proportional to the physical volume of the Bianchi I universe (which has compact $T^3$ topology), with a sign that indicates the orientation of the triad, while $\lambda_\theta$ can be used to measure the anisotropy. On the other hand, in the case of complete isotropy, when all directions present an equivalent behavior and, in particular, $p_\theta$ equals $p_\perp$, one can easily check from the above formulas that $\lambda_\theta=(v/2)^{1/3}$. This remark will be important to understand the physical interpretation of our approximations later on in this work.

We also note that the variables \eqref{var} do not form a canonical set (not even a set closed under Poisson brackets), since we get the following Poisson algebra
\begin{align}\label{poisson}
\{b,v\}&=\frac2{\hbar}, \quad \{b_\theta,\lambda_\theta\}=\frac2{\hbar}\frac{\lambda_\theta}{v}, \qquad \{\lambda_\theta,v\}=0,\nonumber\\\{b_\theta,v\}&=\frac2{\hbar}, \quad \{b_\theta,b\}=\frac2{v} (b_\theta-b),\quad \{\lambda_\theta,b\}=0.
\end{align}
In the loop quantization of the model, one ``polymerizes" the connection, what means that its components $c_\theta$ and $c_\perp$ have no well-defined operator in the quantum theory. Instead, one considers their holonomies along straight edges whose fiducial length is chosen according to the improved dynamics scheme. These holonomies can be described in terms of $e^{\pm i b_\theta}$ and  $e^{\pm i b}$.
The geometry degrees of freedom of the model are represented as operators on a kinematical Hilbert space that can be constructed as the completion of the linear span of the basis of eigenstates $|v,\lambda_\theta\rangle \equiv |v\rangle \otimes |\lambda_\theta\rangle$ of the triad variables defined in Eq. \eqref{var}, where the labels $v,\lambda_\theta\in\mathbb{R}$. The completion is made with respect to the discrete inner product $\langle v',\lambda'_\theta|v,\lambda_\theta\rangle=\delta_{v',v}\delta_{\lambda'_\theta,\lambda_\theta}$. The corresponding operators $\hat{v}$ and $\hat{\lambda}_\theta$ act by multiplication on the elements of this basis:
\begin{align}\label{conf}
\hat{v} |v,\lambda_\theta\rangle=v  |v,\lambda_\theta\rangle,\qquad \hat{\lambda}_\theta  |v,\lambda_\theta\rangle=\lambda_\theta  |v,\lambda_\theta\rangle.
\end{align}
Clearly, they have a discrete spectrum that runs over the whole real line, since all the considered eigenstates have unit norm with the introduced inner product. On the other hand, the holonomy operators cause the following shifts
\begin{align}\label{hol}
\widehat{e^{\pm ib}} |v,\lambda_\theta\rangle=|v\pm2,\lambda_\theta\rangle,\qquad  \widehat{e^{\pm ib_\theta}} |v,\lambda_\theta\rangle=\left|v\pm2, \lambda_\theta\pm\frac{2\lambda_\theta}{v}\right\rangle,
\end{align}
as one can deduce from the canonical commutation relations  $[ \widehat{e^{\pm ib_{a}}},\widehat{v}]=i\hbar\widehat{\{e^{\pm ib_{a}},v}\}$ and $[ \widehat{e^{\pm ib_{a}}},\widehat{\lambda_\theta}]=i\hbar\widehat{\{e^{\pm ib_{a}},\lambda_\theta}\}$, where $b_{a}$ stands both for $b_\theta$ and $b$.

As a result of the loop representation, the classical expressions $ c_\theta p_\theta$ and $c_\perp p_\perp$ get promoted respectively in the quantum theory to the operators \cite{awe1}
\begin{align}
:\!\widehat{\sin(\bar\mu_\theta c_\theta)}\widehat{\left[\frac{p_\theta}{\bar\mu_\theta}\right]}\!\!:\,\,= 2\pi G\hbar \gamma :\!\widehat{v\sin(b_\theta)}\!:\,, \qquad :\!\widehat{\sin(\bar\mu_\perp c_\perp)}\widehat{\left[\frac{p_\perp}{\bar\mu_\perp}\right]}\!\!:\,\,= 2\pi G\hbar \gamma :\!\widehat{v\sin(b)}\!:\,,
\end{align}
where the double dots denote a geometric symmetrization in the factor ordering. More specifically, we choose an ordering analog to that proposed in Refs. \cite{mbmmp,mmo}:
\begin{align}\label{Omegaop}
&2:\!\widehat{v\sin(b)}\!:\,\,\equiv \hat{\Omega}=\sqrt{|\hat v|}\left[\widehat{{\rm sign}(v)}\widehat{\sin(b)}+\widehat{\sin(b)}\widehat{{\rm sign}(v)}\right]\sqrt{|\hat v|},\\
&2:\!\widehat{v\sin(b_\theta)}\!:\,\,\equiv \hat{\Theta}_\theta=\sqrt{|\hat v|}\left[\widehat{{\rm sign}(v)}\widehat{\sin(b_\theta)}+\widehat{\sin(b_\theta)}\widehat{{\rm sign}(v)}\right]\sqrt{|\hat v|}. \label{Thetaop}
\end{align}

The operator $\hat{\Omega}^2$ is not other but the gravitational part of the Hamiltonian constraint for the (well-known) loop quantized flat FRW model \cite{aps, acs, mmo, mop}. As extensively discussed in Refs. \cite{mmo,hybrid4}, with the chosen ordering, states which possess positive $v$ and $\lambda_\theta$ do not get mixed under the action of both  $\hat{\Omega}$ and $\hat{\Theta}_\theta$ with states for which those quantum numbers are negative. Moreover, states with vanishing $v$ and/or $\lambda_\theta$ get totally decoupled. This allows us to remove the latter from the kinematical Hilbert space and, furthermore, to restrict then all considerations e.g. to the subspace spanned by states $|v,\lambda_\theta\rangle$ with $v,\lambda_\theta\in\mathbb{R}^+$. For convenience, instead of working with $\lambda_\theta$ from now on, we will work with its natural logarithm $\Lambda\equiv \log({\lambda_\theta})$.

Finally, adopting a standard Schr\"odinger representation for the homogeneous massless scalar field, $\hat{p}_\phi=-i\hbar\partial_\phi$, the Hamiltonian constraint of the LRS Bianchi I model, given in Eq. \eqref{chi}, is promoted to the symmetric quantum operator
\begin{align}
\hat{C}_{\text{BI}}&=-\frac{\pi G\hbar^2}{8}\left[ 3\hat{\Omega}^2+(\hat{\Omega}\hat{\Theta}+\hat{\Theta}\hat{\Omega})\right] -\frac{\hbar^2\partial_\phi^2}{2}\equiv \hat{C}_{\text{FRW}}-\frac{\pi G\hbar^2}{8}(\hat{\Omega}\hat{\Theta}+\hat{\Theta}\hat{\Omega}).
\end{align}
Here we have introduced the definition $\hat{\Theta}\equiv \hat{\Theta}_\theta-\hat{\Omega}$ with the aim at writing the LRS Bianchi I constraint as that of the flat FRW model coupled to a massless scalar, $\hat{C}_{\text{FRW}}$, plus a term accounting for the anisotropies.

Taking into account the action that $\hat{C}_{\text{BI}}$ has on a generic state $|v,\Lambda \rangle\otimes|\phi\rangle$, which can be easily worked out from the action of the operators defined in Eqs. \eqref{conf} and \eqref{hol}, the geometric sector of the kinematical Hilbert space (with definite sign of $v$ and real $\Lambda$) can be further superselected in separable subspaces. On the one hand, as it happens to be the case in the FRW model, $\hat{C}_{\text{BI}}$ preserves the subspace of states whose quantum number $v$ belongs to any of the semilattices of step four
\begin{align}
\mathcal
L_{\varepsilon}^+=\{\varepsilon+4k;\;k\in\mathbb{N}\},\qquad\varepsilon\in(0,4],
\end{align}
each of them characterized by the minimum value allowed for $v$, namely, the label $\varepsilon$.
On the other hand, concerning the anisotropy variable $\Lambda$, and as discussed in Ref. \cite{hybrid4}, we get that the iterative action of the constraint operator relates any given state $|v,\Lambda^\star\rangle$ only with states whose quantum number $\Lambda$ is of the form
$\Lambda=\Lambda^\star+\omega_{\varepsilon}$, with $\omega_{\varepsilon}$
belonging to the set
\begin{equation}\label{W-set}
\mathcal W_{\varepsilon}=\left\{z\log\left(\frac{\varepsilon-2}{\varepsilon}\right)+\sum_{m,
n\in\mathbb{N}}{k_n^m}\log\left(\frac{\varepsilon+2m}{\varepsilon+2n}\right);\;
k_n^m\in\mathbb{N},\; z\in\mathbb{Z}\text{ if } \varepsilon>2,\;z=0\text{ if }
\varepsilon\le2\right\}.
\end{equation}
This set is countable and dense in the real line \cite{hybrid4}. We will denote by  $\mathcal{S}_\varepsilon\otimes\mathcal{S}^{\varepsilon}_{\Lambda^\star}\otimes\mathcal{S}_\phi$ the domain of definition of  $\hat{C}_{\text{BI}}$, with  $\mathcal{S}_\varepsilon\equiv{\rm span}\{|v\rangle;\, v\in\mathcal{L}_\varepsilon^+\}$, $\mathcal{S}^{\varepsilon}_{\Lambda^\star}\equiv{\rm span}\{|\Lambda^\star+\omega_{\varepsilon}\rangle;\, \omega_{\varepsilon}\in \mathcal{W}_{\varepsilon}\}$, and $\mathcal{S}_\phi$ the Schwartz space of rapidly decreasing functions of $\phi$.
In turn, we denote the corresponding Hilbert space as $\mathcal{H}_\varepsilon\otimes\mathcal{H}^\varepsilon_{\Lambda^\star}\otimes\mathcal{H}_\phi$, where $\mathcal{H}_\varepsilon$ and $\mathcal{H}^\varepsilon_{\Lambda^\star}$ are respectively the completion of $\mathcal{S}_\varepsilon$ and $\mathcal{S}^{\varepsilon}_{\Lambda^\star}$ with respect to the discrete inner product, and $\mathcal{H}_\phi=L^2(\mathbb{R},d\phi)$.

\subsection{Fock quantization of the inhomogeneous sector}

For the  totally deparametrized $T^3$ Gowdy model, it has been shown that there exits a privileged Fock representation for the gravitational and matter fields \cite{cm1,ccm1,ccmv1,cmv}. This Fock representation is the unique one (up to unitary equivalence) admitting a unitary implementation of the dynamics and whose corresponding vacuum is invariant under $S^1$ translations (i.e., rotations in $\theta$), which is the gauge group of the reduced system. The Fock quantization picked out in this way is characterized by a specific choice of configuration variables $\xi$ for the gravitational field  and $\varphi$ for the matter field, choice that in particular involves a time dependent rescaling of the minimally coupled matter scalar field in terms of the homogeneous variables of the Bianchi I sector ($\varphi=\Phi|p_\theta|^{1/2}$). Besides, the selection of canonical momenta for $\xi$ and $\varphi$ is also completely specified by the criteria of unitarity of the dynamics and symmetry invariance that determine the Fock quantization. The two fields $\xi$ and $\varphi$ have the same dynamical behavior and a similar description, since they contribute to the system in the very same way (they have an identical equation of motion, namely, that of a free scalar field with time-dependent mass evolving in a 1+1 Minkowski spacetime). The Fock representation can be constructed by considering as creation and annihilation-like variables those that one would naturally introduce if $\xi$ and $\varphi$ were massless free fields (with canonical momenta coinciding with their times derivatives).

Based on this uniqueness result, we adopt the commented field parametrization $\xi$ and $\varphi$ of Ref. \cite{ccm1} for the gravitational and matter inhomogeneities (i.e., the non-zero modes) of our model, and describe them in terms of the creation and annihilation-like variables mentioned above. By adhering to this Fock quantization we will ensure that, in those regimes where the deparametrized description is valid, the corresponding quantum evolution of the fields is unitary and the gauge symmetry is respected. Thus, we promote our variables to creation and annihilation operators $(\hat{a}^{(\alpha)\dagger}_m,\hat{a}^{(\alpha)}_m)$, with $m\in\mathbb{Z}-\{0\}$ and $\alpha=\xi,\varphi$. We call $\mathcal{F}_\xi\otimes\mathcal{F}_\varphi$ the associated Fock space \cite{wald2}. These operators are densely defined in the space of $n$-particle states $\mathcal{N}_\xi\otimes\mathcal{N}_\varphi\equiv{\rm span}\{|\mathfrak{n}^\xi,\mathfrak{n}^\varphi\rangle\}$, with $\mathfrak{n}^\alpha=\{\cdots, n^\alpha_{-m}, \cdots, n^\alpha_{-1},n^\alpha_{1},\cdots,n^\alpha_{m},\cdots\}$, and where $n^\alpha_{m}\in\mathbb{N}$ is the occupation number in the mode $m$ of the field $\alpha$.

With this representation at hand, we can now promote the momentum constraint (generating rotations in $\theta$) to an operator. It reads
\begin{align}\label{dif}
\hat{C}_\theta=\sum_{\alpha\in{\xi,\varphi}}\,\sum_{m\in\mathbb{N}^{+}} m\left( \hat{a}^{(\alpha)\dagger}_m \hat{a}^{(\alpha)}_m - \hat{a}^{(\alpha)\dagger}_{-m} \hat{a}^{(\alpha)}_{-m}\right).
\end{align}
It only acts on the inhomogeneous sector, imposing a mild restriction on the allowed occupation numbers of the n-particle states $|\mathfrak{n}^\xi,\mathfrak{n}^\varphi\rangle$ (the sum of the occupation numbers multiplied by their respective mode numbers must be zero).

\subsection{Hamiltonian constraint of the quantum hybrid Gowdy model}

With both sectors of the model already represented, we can now promote the Hamiltonian constraint of our Gowdy model to a symmetric operator. The result is \cite{hybrid3,hybrid5}
\begin{align}\label{constra}
\hat{C}_\text{G}=\hat{C}_{\text{FRW}}-\frac{\pi G\hbar^2}{8}(\hat{\Omega}\hat{\Theta}+\hat{\Theta}\hat{\Omega})+
\frac{2\pi G \hbar^{2}}{\beta} \widehat{e^{2\Lambda}}\hat{H}_0+ \frac{\pi G \hbar^2 \beta}{4}  \widehat{e^{-2\Lambda}}\hat{D}\hat{\Omega}^2\hat{D}\hat{H}_\text{I}.
\end{align}
Here $\beta\equiv[G\hbar/(16\pi^{2} \gamma^{2} \Delta)]^{1/3}$.
The last two terms couple the homogeneous background with the inhomogeneities, and contain respectively the free contribution of the non-zero modes
\begin{align}
\hat{H}_0=\sum_{\alpha\in{\xi,\varphi}}\,\sum_{m\in\mathbb{Z}-\{0\}} |m| \, \hat{a}^{(\alpha)\dagger}_m \hat{a}^{(\alpha)}_m,
\end{align}
a self-interaction term for these inhomogeneities
\begin{align}
\hat{H}_\text{I}=\sum_{\alpha\in{\xi,\varphi}}\,\sum_{m\in\mathbb{Z}-\{0\}}\frac1{2|m|} \left(2 \hat{a}^{(\alpha)\dagger}_m \hat{a}^{(\alpha)}_m+\hat{a}^{(\alpha)\dagger}_m \hat{a}^{(\alpha)\dagger}_{-m}+\hat{a}^{(\alpha)}_m \hat{a}^{(\alpha)}_{-m}\right),
\end{align}
and a term coming from a regularization of the inverse of the volume  \cite{hybrid3}:
\begin{align}
\hat{D}|v\rangle=D(v)|v\rangle,\qquad D(v)\equiv v\left(\sqrt{v+1}-\sqrt{|v-1|}\right)^2.
\end{align}
For large values of $v$, the function $D(v)$ approaches the unit, whereas it vanishes in the limit where $v$ tends to zero. In the construction of the free and self-interaction terms for the inhomogeneities, we have adopted normal ordering with respect to the vacuum of the selected representation (namely, the one which would be natural in the massless case)\footnote{We note that other choices of ordering corresponding to other physically interesting vacua, in the unitary equivalence class picked out by the criteria of invariance and unitary dynamics, would lead to similar results. For instance, one might adopt normal ordering with respect to some adiabatic vacuum state. The corresponding representation is in the unitary equivalence class that we are considering, as can be proven by arguments likes those discussed in Ref.~\cite{desitter}. Moreover, it is possible to see that, in the Gowdy model, the difference in zero point energy for this alternate choice would be finite.}.

Our constraint operator is well defined in the Hilbert space $\mathcal{H}_\varepsilon\otimes\mathcal{H}^\varepsilon_{\Lambda^\star}\otimes\mathcal{H}_\phi\otimes\mathcal{F}_\xi\otimes\mathcal{F}_\varphi$, with dense domain given by the subspace $\mathcal{S}_\varepsilon\otimes\mathcal{S}^{\varepsilon}_{\Lambda^\star}\otimes\mathcal{S}_\phi\otimes\mathcal{N}_\xi\otimes\mathcal{N}_\varphi$. We see that, if we ignore the inhomogeneities and the anisotropies, the Hamiltonian constraint of the Gowdy model reduces to that of  the flat FRW coupled to a homogeneous massless scalar field, $\hat{C}_{\text{FRW}}$. We will focus our attention on a regime where the effects of the anisotropies and the interactions between the inhomogeneities can be disregarded, so that then the Gowdy model can be viewed, in the commented sense, as a flat FRW background with gravitational and matter ``perturbations'' coupled between them only by the addition of their free-field energy contribution to the Hamiltonian constraint [as well as by the constraint \eqref{dif} of vanishing total field momentum].

\section{Approximating the quantum Hamiltonian constraint}
\label{sec:App}

The quantum Hamiltonian constraint $\hat{C}_\text{G}$ is excessively complicated to find its solutions, $(\Psi|\hat{C}_{\text{G}}^{\dagger}=0$ (here, the dagger denotes the adjoint). Even assuming that $\hat{C}_\text{G}$ is essentially self-adjoint, its intricate expression prevents us from determining the form of the (generalized) states in its kernel. This obstruction is due mainly to three facts. First, regarding the operators that affect the gravitational part of the homogeneous sector, we have that $\hat{D}$ and $\hat{\Omega}^2$ do not commute, and hence cannot be diagonalized simultaneously, complicating the resolution in the variable $v$. Besides, and more importantly, $\hat{\Omega}^2$ and $\hat{\Omega}\hat{\Theta}+\hat{\Theta}\hat{\Omega}$, defined on $\mathcal{S}_\varepsilon\otimes\mathcal{S}^{\varepsilon}_{\Lambda^\star}$, do not commute either. This extends the problems for solving the constraint, as far as the homogeneous sector is involved, to the whole of the geometry degrees of freedom. Finally, while the operator $\hat{H}_0$ acts diagonally on the $n$-particle states, the operator $\hat{H}_\text{I}$ creates and annihilates a pair of particles in every mode, and therefore creates an infinite number of them (nonetheless, let us emphasize that it is well defined in $\mathcal{N}_\xi\otimes\mathcal{N}_\varphi$ \cite{hybrid3}). These complications call for the introduction of approximations in order to solve the Hamiltonian constraint, both for the Gowdy model and for the particular case in which the inhomogeneities are unplugged.

For the regime of negligible effects of the anisotropies and of the inhomogeneities interactions that we are interested in, it is convenient to diagonalize the operator acting on the homogeneous and isotropic sector, and use a basis of its eigenstates for our discussion, namely, a basis of the operator
\begin{align}
\hat{C}_{\text{FRW}}&=-\frac{3\pi G\hbar^2}{8}\hat{\Omega}^2-\frac{\hbar^2\partial_\phi^2}{2}.
\end{align}

\subsection{Spectral analysis of $\hat{C}_{\text{FRW}}$}

As it is well known, the operator $-\hbar^2\partial_\phi^2$ in $\mathcal{S}_\phi$ is essentially self-adjoint, and has an absolutely continuous, doubly-degenerate, and positive spectrum. Its delta-normalized (generalized) eigenstates with eigenvalue $p_\phi^2$ are the plane waves $|e_{\pm p_\phi}\rangle$, with wavefunction
\begin{align}
e_{p_\phi}(\phi)\equiv\langle \phi |e_{p_\phi}\rangle=\frac1{\sqrt{2\pi\hbar}}e^{\frac{i}{\hbar}p_\phi\phi},\qquad \langle e_{p_\phi'}|e_{p_\phi}\rangle=\delta({p_\phi}'-{p_\phi}).
\end{align}
They provide a resolution of the identity in $\mathcal{H}_\phi$ given by $\mathbb{I}_{\mathcal{H}_\phi}=\int_{-\infty}^{\infty} dp_\phi |e_{ p_\phi}\rangle \langle e_{p_\phi}|$.

On the other hand, the operator $\hat{\Omega}^2$ has been analyzed in detail in the LQC literature (see e.g. Refs.~\cite{lqc2,hom4,mmo}). It is essentially self-adjoint with domain in $\mathcal{S}_\varepsilon$, and has an absolutely continuous, non-degenerate, and positive spectrum. In each of the superselection sectors corresponding to the semilattices $\mathcal{L}^+_\varepsilon$, its delta-normalized eigenstates with eigenvalue $\rho^2$, $|e^\varepsilon_\rho\rangle=\sum_{v\in\mathcal{L}^+_\varepsilon}e^\varepsilon_\rho(v)|v\rangle$, are defined via the recurrence relations
\begin{align}
e^\varepsilon_\rho(\varepsilon+4)&=\frac{2\varepsilon^2-\rho^2}{(\varepsilon+2)\sqrt{\varepsilon(\varepsilon+4})}e^\varepsilon_\rho(\varepsilon),\nonumber\\
e^\varepsilon_\rho(v+4)&=\frac{2v^2-\rho^2}{(v+2)\sqrt{v(v+4)}}e^\varepsilon_\rho(v)-\frac{(v-2)\sqrt{v(v-4)}}{(v+2)\sqrt{v(v+4)}}e^\varepsilon_\rho(v-4), \quad \text{if } v=\varepsilon+4n>4,
\end{align}
such that $ \langle e^\varepsilon_{\rho'}|e^\varepsilon_\rho\rangle=\delta({\rho'}-{\rho})$. Note that the eigenfunction $e^\varepsilon_\rho(v)$ is then real, up to a constant phase, determined by $e^\varepsilon_\rho(\varepsilon)$. These generalized eigenstates provide a resolution of the identity in $\mathcal{H}_\varepsilon$, given by $\mathbb{I}_{\mathcal{H}_\varepsilon}=\int_{0}^{\infty} d\rho |e^\varepsilon_{ \rho}\rangle \langle e^\varepsilon_\rho|$.

\begin{figure}[ht]
 \includegraphics[width=0.48\textwidth]{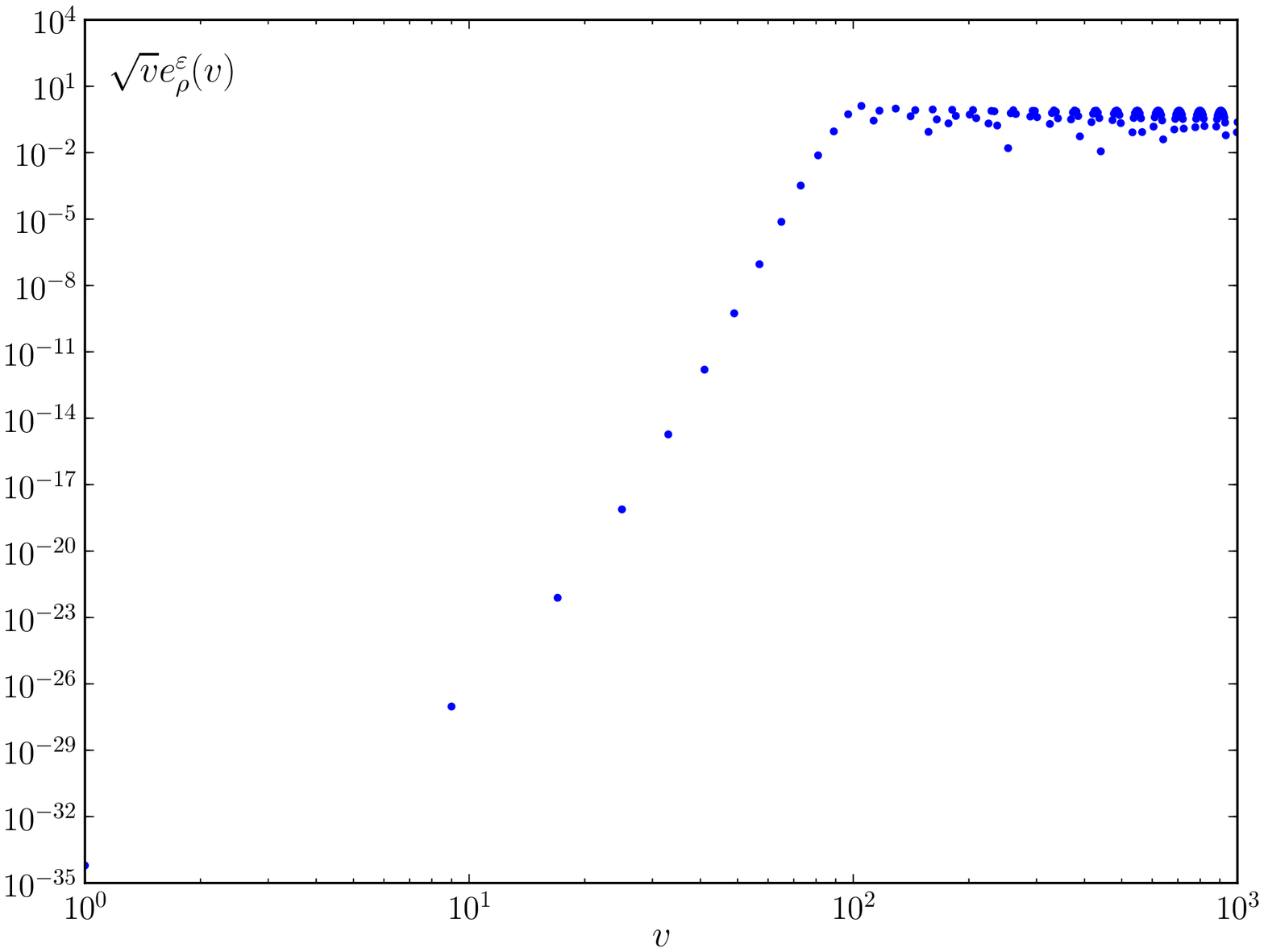}
 \includegraphics[width=0.48\textwidth]{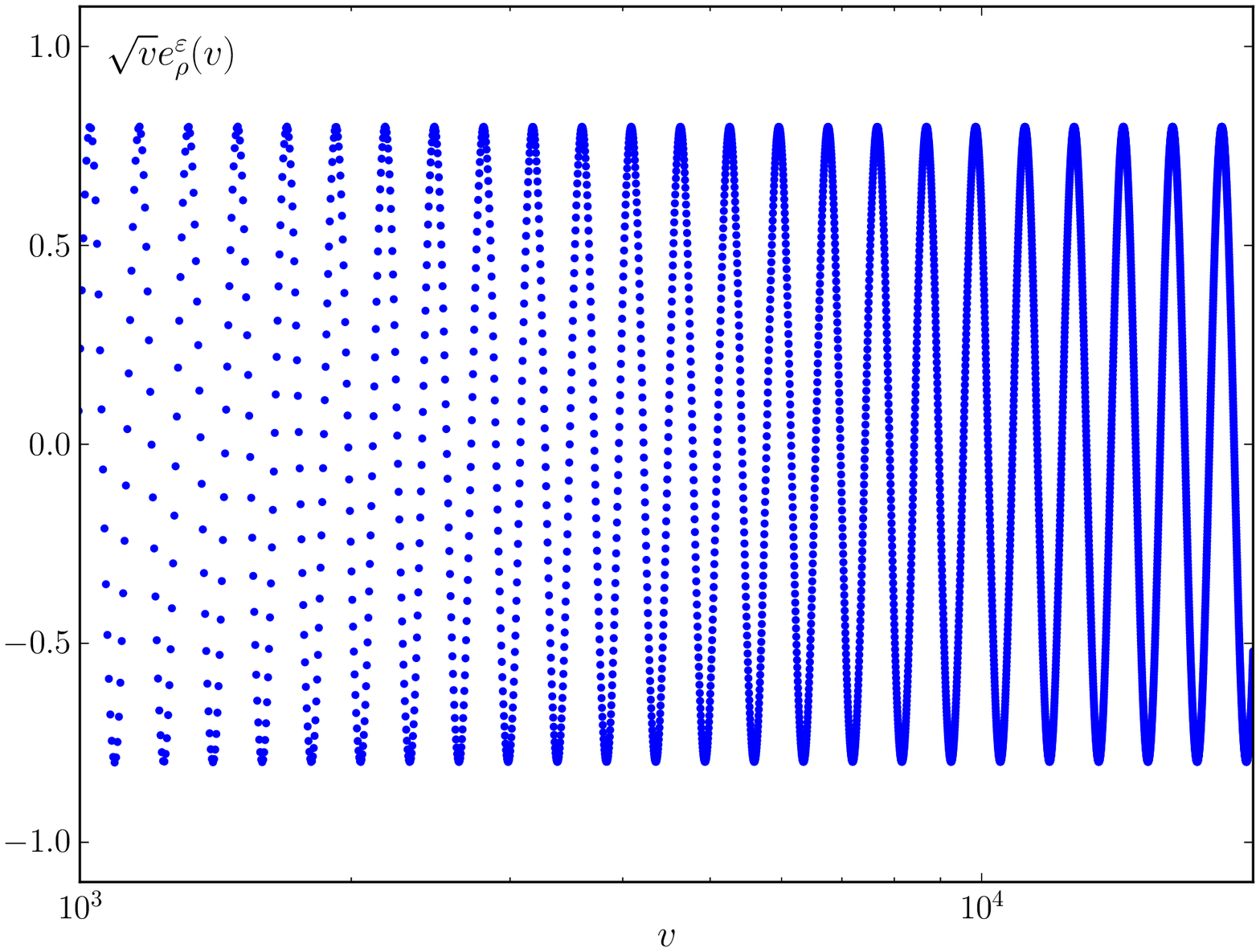}
\caption{Function  $\sqrt{v}e^\varepsilon_\rho(v)$ for $\varepsilon=1$ and $\rho=200$. The graph on the left shows the exponential suppression for small values of $v$, whereas the graph on the right displays the oscillatory behavior at large $v$. A logarithmic scale is used in the horizontal axis of both graphs.}
\label{fig:FRWaut}
\end{figure}

We have computed these eigenfunctions numerically. They have the following properties:
\begin{itemize}
\item $e^\varepsilon_\rho(v)$ is exponentially suppressed for $v\lesssim v_{\text{min}}(\rho)\equiv \rho/2$ (see left graph in Fig. \ref{fig:FRWaut}).
\item For $v\gg v_{\text{min}}(\rho)$, these eigenfunctions converge to a linear combination of two eigenfunctions of the operator analog to $\hat{\Omega}^2$ in the Wheeler-de Witt (WDW) quantization \cite{mmo}. This WDW limit is given by the function
\begin{align}
e^\varepsilon_\rho(v)\simeq \underline{e}^\varepsilon_\rho(v)\equiv\sqrt{\frac{2}{\pi v}}\cos\left(\frac{\rho}{4}\log(v) +\phi^\varepsilon_{\rho}\right), \quad \phi^\varepsilon_{\rho}=-\frac{\pi\varepsilon}{4}+R_\varepsilon(\rho)+T(\rho),
\end{align}
with $\lim_{\rho\rightarrow\infty}R_\varepsilon(\rho)=0$ and $T(\rho)$ a function of $\rho$ independent of $\varepsilon$ \cite{mop,kpa}. This oscillating behavior can be seen very clearly in the example shown in Fig.~\ref{fig:FRWaut}.
\end{itemize}

When studying the solutions of the Hamiltonian constraint of the Gowdy model, we will expand the homogeneous and isotropic part of our states $|\Psi\rangle$ in the basis provided by the generalized eigenstates of the FRW constraint operator, $|e^\varepsilon_\rho\rangle\otimes|e_{p_\phi}\rangle$, so that at least the action of $\hat{C}_{\text{FRW}}$ becomes diagonal, and we can easily find out the relation between the analyzed (approximate) solutions for the Gowdy cosmologies and those of the FRW model. Still, the Hamiltonian constraint, given in Eq. \eqref{constra}, presents non-diagonal contributions acting non-trivially on this homogeneous and isotropic sector: the term involving the operator $\hat{D}$, and the anisotropy term $\hat{\Omega}\hat{\Theta}+\hat{\Theta}\hat{\Omega}$. In the following, we explain how to approximate these terms by much more manageable operators.

\subsection{Approximation 1: Dealing with the inverse volume corrections}

\begin{figure}[ht]
  \includegraphics[width=0.48\textwidth]{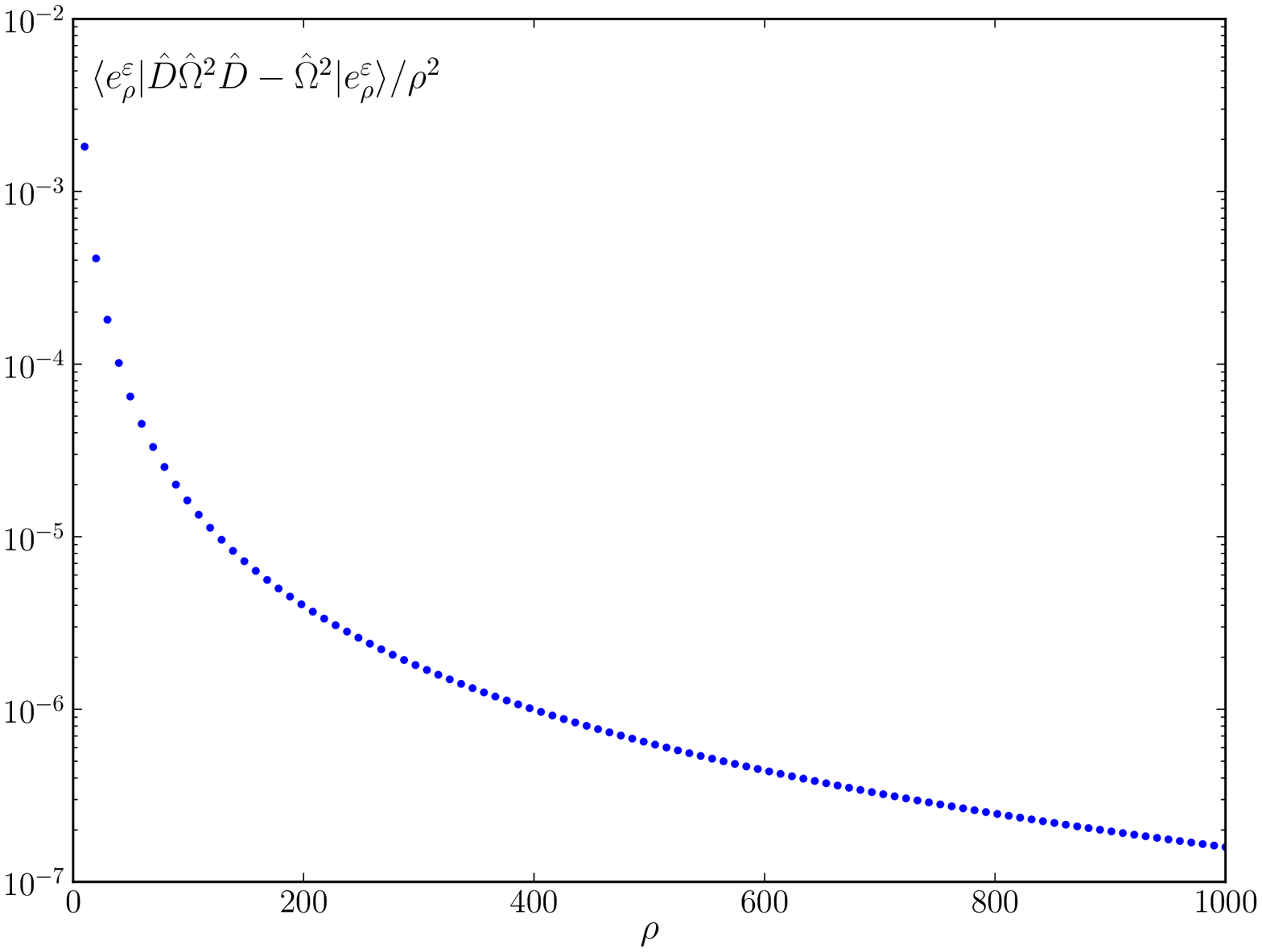}
  \includegraphics[width=0.48\textwidth]{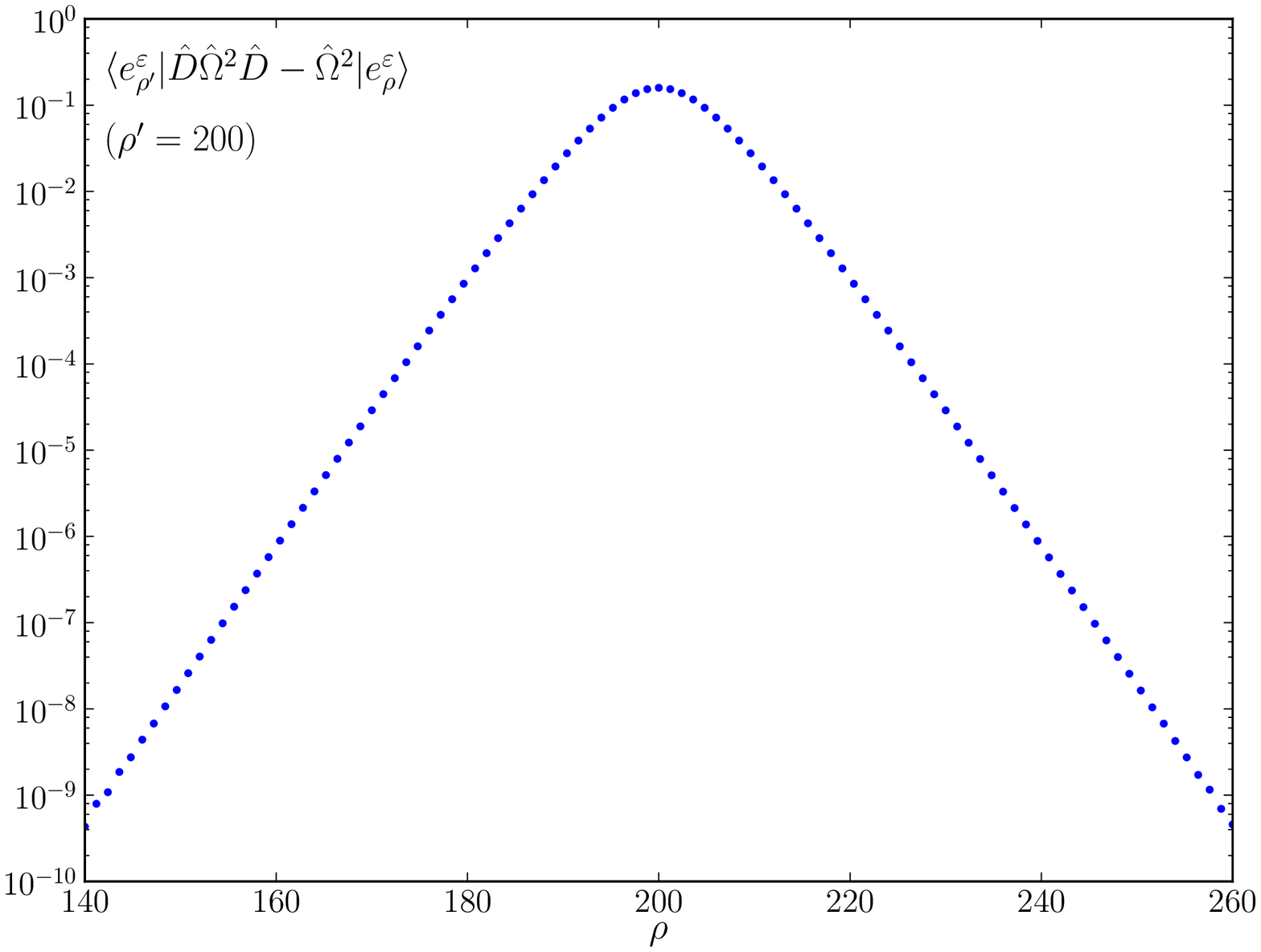}
\caption{Difference between the matrix elements of the operators $\hat{D}\hat{\Omega}^{2}\hat{D}$ and $\hat{\Omega}^{2}$. The graph on the left shows the difference between the diagonal matrix elements, relative to the eigenvalue $\rho^{2}$, whereas the graph on the right shows the difference between non-diagonal matrix elements with $\rho'=200$. Sums are truncated at $v=4800001$.}
\label{fig:DOOD}
\end{figure}

The behavior of the eigenfunctions $e^\varepsilon_\rho(v)$ allows us to deal with the first complication that we have mentioned, concerning the fact that $\hat{\Omega}^2$ and $\hat{D}$ do not commute. Indeed, for $\rho/8$ much larger than the unit, let's say $\rho\gg 10$, we have
\begin{align}
\hat{D}\hat{\Omega}^2\hat{D} |e^\varepsilon_\rho\rangle \simeq\rho^2   |e^\varepsilon_\rho\rangle=\hat{\Omega}^2 |e^\varepsilon_\rho\rangle.
\end{align}
This result was expected since $D(v)\simeq 1$ for $v\gg 1$, and the eigenfunctions of the gravitational part of the FRW constraint operator are exponentially negligible for $v< \rho/2$. We have checked this approximation by comparing the matrix elements $\langle e^\varepsilon_{\rho'}| \hat{D}\hat{\Omega}^2\hat{D} |e^\varepsilon_\rho\rangle$ and $\langle e^\varepsilon_{\rho'}| \hat{\Omega}^2 |e^\varepsilon_\rho\rangle$. This comparison has been performed:

i) analytically, by approximating $e^\varepsilon_\rho(v)$ by its WDW limit  $\underline{e}^\varepsilon_\rho(v)$ for $v>\rho/2$ and by zero for $v\leq\rho/2$, and estimating then the discrete sums in $v$ by means of integral expressions, making the replacement \begin{align}
\sum_{(\rho_M/2)<v\in\mathcal{L}_\varepsilon^+} \;\longrightarrow\; \frac1{4}\int_{\rho_M/2}^\infty \!\!\!dv \ \ ,
\end{align}
where $\rho_M$ is the largest of $\rho$ and $\rho'$;

ii) numerically, by computing the two considered types of matrix elements, namely, those of the operators $\hat{D}\hat{\Omega}^2\hat{D}$ and $\hat{\Omega}^2$. The difference between these two matrix elements, relative to the eigenvalue $\rho^2$, is displayed in Fig. \ref{fig:DOOD} for $\rho$ ranging between $10$ and $1000$. The graph on the left shows the diagonal matrix elements, whereas the graph on the right corresponds to the non-diagonal ones with fixed $\rho'=200$.

It is worth mentioning that the numerical computation of these diagonal matrix elements separately would diverge if all the infinite number of contributions in the sum over the variable $v\in\mathcal{L}^{+}_{\varepsilon}$ were included. This would happen for instance for $\langle e^{\varepsilon}_{\rho}|\hat{\Omega}^{2}|e^{\varepsilon}_{\rho}\rangle$. Nonetheless, numerically, the difference between the matrix elements of the two operators remains finite even in this limit in which the cumulative sum is extended to infinite large values of $v$.

Hence, for eigenstates $|e^\varepsilon_\rho\rangle$ with $\rho \gg 10$, the contributions of the $\hat{D}$ operators in the interaction term of the Hamiltonian constraint \eqref{constra} can be disregarded.

\subsection{Approximation 2: Dealing with the anisotropy terms}

The anisotropy term $\hat{\Omega}\hat{\Theta}+\hat{\Theta}\hat{\Omega}$ is more difficult to handle. To start with, it does not factorize in two operators acting on $\mathcal{H}_\varepsilon$ and $\mathcal{H}^\varepsilon_{\Lambda^\star}$ separately, but acts on $\mathcal{H}_\varepsilon\otimes\mathcal{H}^\varepsilon_{\Lambda^\star}$ as a whole. On the other hand, the action upon the anisotropy label ${\Lambda}$ is quite complicated, something that makes a spectral analysis of the operator unviable in practice. To overcome these problems we will approximate the operator $\hat{\Omega}\hat{\Theta}+\hat{\Theta}\hat{\Omega}$ by another operator $2\hat{\Omega}'\hat{\Theta}'$ such that: i) it factorizes, namely $\hat{\Omega}'$ is well defined in $\mathcal{H}_\varepsilon$ (with domain of definition $\mathcal{S}_\varepsilon$) and $\hat{\Theta}'$ is well defined in  $\mathcal{H}^\varepsilon_{\Lambda^\star}$ (with domain of definition $\mathcal{S}^\varepsilon_{\Lambda^\star}$); and ii) $\hat{\Theta}'$ produces a constant shift in the anisotropy label, so that its domain can be further restricted in the superselection sector just to states with support in a lattice of constant step.

Let us summarize our approach to get that approximation. First of all, we consider the action of $\hat{\Omega}\hat{\Theta}+\hat{\Theta}\hat{\Omega}$ on states $|e_{\rho}^{\varepsilon}\rangle \otimes | f(\Lambda)\rangle$, with $|f(\Lambda)\rangle=\sum_{\omega_{\varepsilon}\in\mathcal{W}_\varepsilon}f(\Lambda^\star+\omega_{\varepsilon})|\Lambda^\star+\omega_{\varepsilon}\rangle$. We restrict our study exclusively to states for which $f(\Lambda)$ can be extended to a smooth function in the real line that, for variations in $\Lambda$ smaller than a certain scale $q_{\varepsilon}$ (possibly depending on the considered superselection sector $\mathcal{H}^\varepsilon_{\Lambda^\star}$), satisfies:
\begin{equation}\label{tayl}
f(\Lambda+\Lambda_{0})\simeq f(\Lambda) +\Lambda_{0}\partial_{\Lambda} f(\Lambda).
\end{equation}
Thus, for $\Lambda_0\leq q_{\varepsilon}$, our approximation disregards higher-order terms in $\partial_\Lambda$ in a possible Taylor expansion. Let us notice that, since contributions with $v < \rho/2$ are exponentially suppressed in $|e_{\rho}^{\varepsilon}\rangle$, the shift that $\hat{\Theta}$ produces in $\Lambda$ (with a non-negligible contribution) is never greater than $\log{(1+4/\rho)}$ for the states that we are studying. Therefore, if we are considering states that only have relevant FRW contributions with $\rho\geq \rho^{\star}$, we expect our approximation to be valid for anisotropy wavefunctions that are smooth and do not change much at scales $q_{\varepsilon}\geq \log{(1+4/\rho^{\star})}$. Then, under these conditions, we get that $\hat{\Omega}\hat{\Theta}+\hat{\Theta}\hat{\Omega}$ can be approximated by the operator $-8i\hat{\Omega}' \partial_\Lambda$, with
\begin{align}\label{omegaprime}
\hat{\Omega}' |v\rangle=\frac{i}{8}\left[\log\left(1+\frac{4}{v}\right)y_{++}(v)|v+4\rangle+\log\left(1-\frac{4}{v}\right)y_{--}(v)|v-4\rangle\right],
\end{align}
and
\begin{align}\label{def2}
y_{\pm\pm}(v)=y_\pm(v) y_\pm(v\pm2),\qquad y_\pm(v)=\frac{1+\text{sign}(v\pm2)}{2}\sqrt{v(v\pm2)}.
\end{align}
The operator $\hat{\Omega}'$ is well defined in any of the superselection sectors $\mathcal{H}_\varepsilon$.

The details of the derivation of this approximation are as follows. As it can be easily deduced from the definitions \eqref{Omegaop} and \eqref{Thetaop}, the action of the operators $\hat{\Omega}$ and $\hat{\Theta}$ is
\begin{align}
\hat{\Omega}|v\rangle= i \left[y_-(v)|v-2\rangle - y_+(v)|v+2\rangle\right],
\end{align}
\begin{align}
\hat{\Theta}|v\rangle\otimes|\Lambda\rangle&= i[ y_-(v)|v-2\rangle\otimes\left(|\Lambda+d_v(-2)\rangle-|\Lambda\rangle\right)-y_+(v)|v+2\rangle\otimes\left(|\Lambda+d_v(2)\rangle-|\Lambda\rangle\right)],
\end{align}
with
\begin{align}\label{def2}
d_v(n)\equiv \log\left(1+\frac{n}{v}\right).
\end{align}
Therefore, the action of the anisotropy term $\hat{\Omega}\hat{\Theta}+\hat{\Theta}\hat{\Omega}$ reads
\begin{align}\label{eq:ota}
(&\hat{\Omega}\hat{\Theta}+\hat{\Theta}\hat{\Omega})|v,\Lambda\rangle  = \nonumber \\
& =  -y_{--}(v)|v-4\rangle\otimes\left(|\Lambda+d_{v}(-4)-d_{v}(-2)\rangle-2|\Lambda\rangle+|\Lambda+d_{v}(-2)\rangle\right) \nonumber \\
&  \quad   + y_{+-}(v)|v\rangle\otimes\left(|\Lambda-d_{v}(-2)\rangle-2|\Lambda\rangle+|\Lambda+d_{v}(-2)\rangle\right) \nonumber\\
& \quad + y_{-+}(v)|v\rangle\otimes\left(|\Lambda-d_{v}(2)\rangle-2|\Lambda\rangle+|\Lambda+d_{v}(2)\rangle\right) \nonumber\\
& \quad - y_{++}(v)|v+4\rangle\otimes\left(|\Lambda+d_{v}(4)-d_{v}(2)\rangle-2|\Lambda\rangle+|\Lambda+d_{v}(2)\rangle\right),
\end{align}
with $y_{\pm\pm}(v)$ as defined above and
\begin{align}\label{defini2}
 y_{\mp\pm}(v)=y_\pm(v) y_\mp(v\pm2).
\end{align}

Acting on states $|v, f(\Lambda)\rangle$ with $|f(\Lambda)\rangle$ of the form that we have explained, and projecting on the anisotropy variable, we get
\begin{align}
\langle\Lambda|(&\hat{\Omega}\hat{\Theta}+\hat{\Theta}\hat{\Omega})|v,f(\Lambda)\rangle  = \nonumber \\
& = -y_{--}(v)|v-4\rangle\left[f(\Lambda-d_{v}(-4)+d_{v}(-2))-2f(\Lambda)+f(\Lambda-d_{v}(-2))\right] \nonumber \\
&  \quad  + y_{+-}(v)|v\rangle\left[f(\Lambda+d_{v}(-2))-2f(\Lambda)+f(\Lambda-d_{v}(-2))\right] \nonumber\\
& \quad + y_{-+}(v)|v\rangle\left[f(\Lambda+d_{v}(2))-2f(\Lambda)+f(\Lambda-d_{v}(2))\right] \nonumber\\
& \quad - y_{++}(v)|v+4\rangle\left[f(\Lambda-d_{v}(4)+d_{v}(2))-2f(\Lambda)+f(\Lambda-d_{v}(2))\right].
\end{align}

For states for which $f(\Lambda)$ can be extended to a sufficiently smooth function in the real line such that the approximation \eqref{tayl} is good for variations $\Lambda_0$ smaller than the scale $q_{\epsilon}$, we obtain
\begin{align}
\langle\Lambda|(&\hat{\Omega}\hat{\Theta}+\hat{\Theta}\hat{\Omega})|v,f(\Lambda)\rangle  \simeq [d_v(-4)y_{--}(v)|v-4\rangle+d_v(4)y_{++}(v)|v+4\rangle]\partial_\Lambda f(\Lambda).
\end{align}
From this, we then see that the operator $\hat{\Omega}\hat{\Theta}+\hat{\Theta}\hat{\Omega}$ acting on the considered states $|v,f(\Lambda)\rangle$ can be approximated by the operator $-8i\hat{\Omega}' \partial_\Lambda$, as we wanted to proof.

At this stage, we can still improve our approximation in two different aspects. First, when taking the approximation \eqref{tayl}, we have ignored the discrete nature of the geometry, extending the wavefunction $f(\Lambda)$ to the real line, instead of treating it as a function of the anisotropy label defined in the superselection sector $\Lambda-\Lambda^{\star}\in \mathcal{W}_\varepsilon$. Now we will restore that discreteness to capture the essence of the loop quantization in the anisotropies. With this aim, we approximate the first derivative $-4i\partial_\Lambda$ by a discrete derivative at the scale $q_{\varepsilon}$
\begin{align}
\hat{\Theta}'|\Lambda^\star\rangle\equiv i\frac2{q_\varepsilon}\left(|\Lambda^\star+q_\varepsilon\rangle-|\Lambda^\star-q_\varepsilon\rangle\right).
\end{align}
Note that $q_\varepsilon$ must be a point in the set $\mathcal{W}_\varepsilon$ in order to preserve the original superselection sector. If, according to our previous discussion, we only consider non-negligible gravitational FRW contributions in the sector $\rho\geq \rho^{\star}$, it suffices to adopt a scale $\mathcal{W}_\varepsilon\ni q_\varepsilon\geq \log{(1+4/\rho^{\star})}$. The shift  $q_\varepsilon$ must be small enough for the approximation to apply on sufficiently general anisotropy wavefunctions, allowing us to disregard contributions coming from higher-order difference terms (i.e., higher-order derivatives in the continuum). Thus, the closest $q_\varepsilon$ is to $\log{(1+4/\rho^{\star})}$, the better is our approximation. Later on, when constructing approximate solutions to the Gowdy Hamiltonian constraint, we will comment further on the choice of this parameter.

The discrete derivative $\hat{\Theta}'$ is well defined in $\mathcal{S}^{q_\varepsilon}_{\Lambda^\star}={\rm span}\{|\Lambda\rangle;\, \Lambda\in\mathcal{L}^{q_\varepsilon}_{\Lambda^\star} \}\subset \mathcal{S}^{\varepsilon}_{\Lambda^\star}$, where $\mathcal{L}^{q_\varepsilon}_{\Lambda^\star} =\{ \Lambda^\star+nq_\varepsilon;\, n\in\mathbb{Z}\}$ is a lattice of constant step. The closure of $\mathcal{S}^{q_\varepsilon}_{\Lambda^\star}$ in the discrete norm will be called  $\mathcal{H}^{q_\varepsilon}_{\Lambda^\star}\subset  \mathcal{H}^\varepsilon_{\Lambda^\star}$.

The second aspect in which we can improve our approximation, at least for practical purposes, concerns the operator $\hat{\Omega}'$. Recall that the contribution of the region $v< \rho/2$ is exponentially suppressed in the FRW eigenfunctions. Therefore, for $\rho \gg 10$, we can restrict all considerations to the sector $v\gg 1$. In this sector, one can approximate $\log(1\pm4/v)(v\pm2)$ by $\pm 4$ and, consequently, functions of $v$ appearing in the definition \eqref{omegaprime} can be approximated so that $\hat{\Omega}'$ becomes
\begin{align}
\hat{\tilde\Omega}=\sqrt{\frac{|\hat v|}{2}}\left[\widehat{{\rm sign}(v)}\widehat{\sin(2b)}+\widehat{\sin(2b)}\widehat{{\rm sign}(v)}\right]
\sqrt{\frac{|\hat v|}{2}},
\end{align}
which is completely analogous to the geometry operator $\hat{\Omega}$, introduced in Eq. \eqref{Omegaop}, except for the replacement of the canonically conjugate variables $(v,b)$ by the new pair $(v/2,2b)$. In particular, this analogy shows immediately that $\hat{\tilde\Omega}$, with the domain  $\mathcal{S}_{\varepsilon}$, is a well-defined operator in $\mathcal{H}_\varepsilon$, and that it is essentially self-adjoint, with an absolutely continuous, non-degenerate spectrum that coincides with the real line.

\begin{figure}[ht]
  \includegraphics[width=0.48\textwidth]{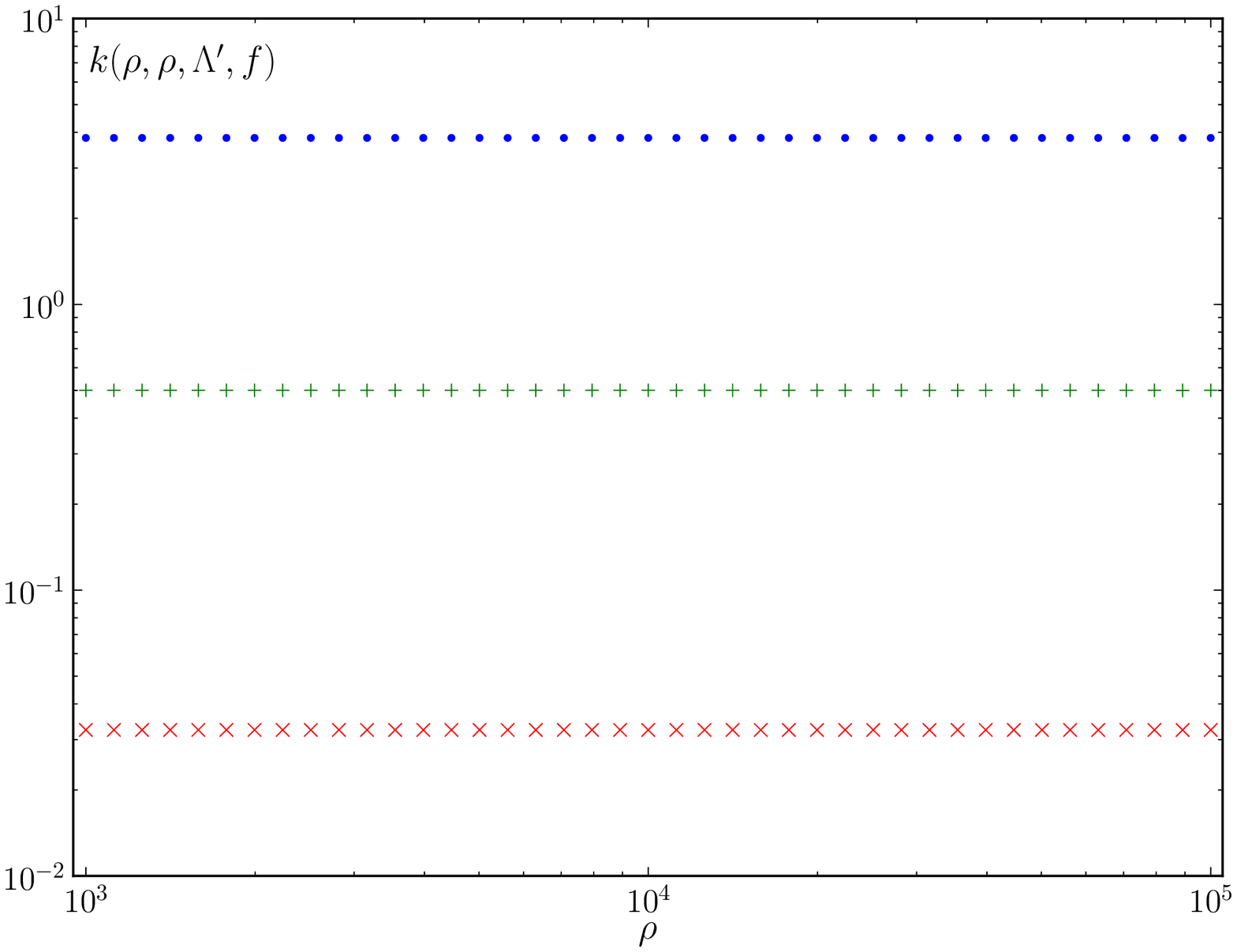}
  \includegraphics[width=0.48\textwidth]{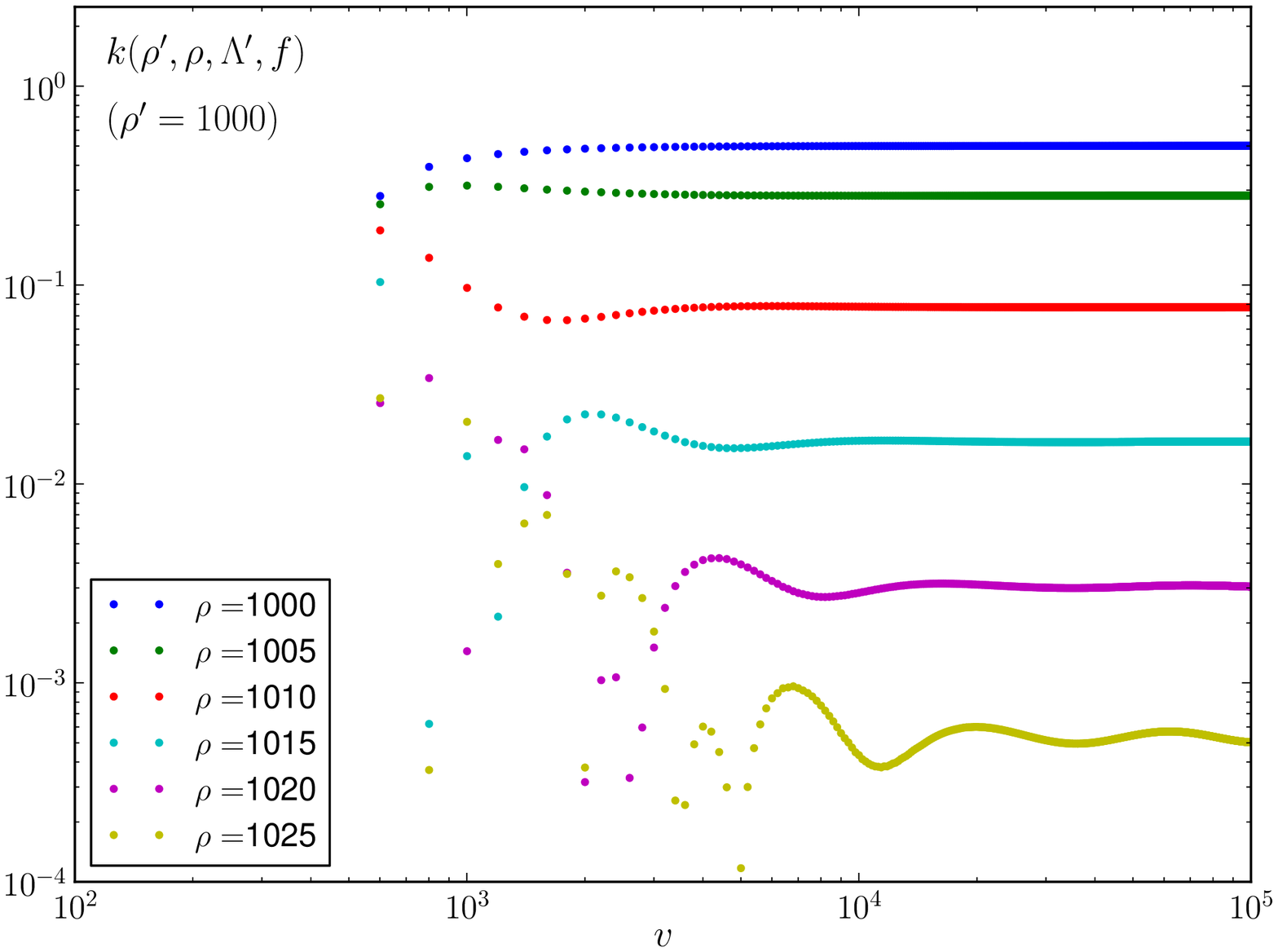}
\caption{Matrix elements of the difference between the two considered anisotropy terms, namely, $\langle\Lambda'|\otimes\langle e^\varepsilon_{\rho'}|\left[2\hat{\tilde\Omega}\hat{\Theta}'-(\hat{\Omega}\hat{\Theta}+\hat{\Theta}\hat{\Omega})\right] \sum_{\Lambda} f(\Lambda) |e^\varepsilon_{\rho}\rangle\otimes|\Lambda\rangle\equiv k(\rho',\rho,\Lambda',f)$, with $\Lambda'=0.1$ and Gaussian profiles centered at $\bar{\Lambda}=0.0$. The graph on the left shows diagonal matrix elements for three different values of the Gaussian profile width: $\sigma_{\Lambda}=0.5$ (blue dots), $\sigma_{\Lambda}=1.0$ (green crosses), and $\sigma_{\Lambda}=2.5$ (red x-shaped crosses). The corresponding sums have been truncated at $v=1200001$. The graph on the right displays the cumulative sums (truncated at different values of $v$) for non-diagonal matrix elements with $\rho'=1000$ and $\sigma_{\Lambda}=1.0$.}
\label{fig:TOOT}
\end{figure}

To check our approximation, we have computed numerically matrix elements of the type $\langle\Lambda'|\otimes\langle e^\varepsilon_{\rho'}| (\hat{\Omega}\hat{\Theta}+\hat{\Theta}\hat{\Omega}) \sum_{\Lambda} f(\Lambda) |e^\varepsilon_{\rho}\rangle\otimes|\Lambda\rangle$ and $\langle\Lambda'|\otimes\langle e^\varepsilon_{\rho'}| 2\hat{\tilde\Omega}\hat{\Theta}' \sum_{\Lambda} f(\Lambda) |e^\varepsilon_{\rho}\rangle\otimes|\Lambda\rangle$ for smooth wavefunctions $f(\Lambda)$, namely, Gaussian profiles of the form
\begin{align}
f(\Lambda)=\frac{1}{\sqrt{2\pi\sigma^{2}_{\Lambda}}}\exp\left[-\frac{\left(\Lambda-\bar{\Lambda}\right)^{2}}{2\sigma^{2}_{\Lambda}}\right].
\end{align}
The comparison demonstrates that in fact the approximation $\hat{\Omega}\hat{\Theta}+\hat{\Theta}\hat{\Omega}\simeq 2\hat{\tilde\Omega}\hat{\Theta}' $ is fairly good for this kind of wavefunctions, provided condition \eqref{tayl} is satisfied. In Fig. \ref{fig:TOOT} we show the difference between these two types of matrix elements {. More specifically, we show the results obtained for $\Lambda'=0.1$ with Gaussian profiles peaked at $\bar{\Lambda}=0$. In the graph on the left the diagonal elements are displayed with $\rho$ ranging between $1000$ and $100000$, and three different values of the Gaussian profile width: $\sigma_{\Lambda}=0.5$, $1.0$, and $2.5$. The graph on the right shows the cumulative sums (truncated at different values of $v$) of the non-diagonal elements for a Gaussian profile width $\sigma_{\Lambda}=1.0$, with $\rho'=1000$ and six different values of $\rho$, equispaced in the interval $[1000,1025]$. For these numerical computations, we have chosen $q_{\varepsilon}=\log(1+2/v_{\rho^{\star}})\in\mathcal{W}_{\varepsilon}$ with $\rho^{\star}=1000$, where $v_{\rho^{\star}}=\max\{v;\,v\in\mathcal{L}^{+}_{\varepsilon},\, v< \rho^{\star}/2\}$.}

The diagonal (i.e., $\rho'=\rho$) matrix elements of the difference between the exact and the approximate operator prove to be finite and (almost) independent of the value of $\rho$ when $\rho \gg 10$. Furthermore, they do not depend on the chosen value of $q_{\varepsilon}$, but only on the studied profile, and more precisely on the value of the derivatives of order higher than one at the considered point $\Lambda'$ (mainly on the second derivative). One can check that these finite values decrease {to zero when those derivatives are negligible, i.e., when condition \eqref{tayl} is satisfied. On the other hand, when increasing the value of $v$ at the truncation, the cumulative sums for the non-diagonal matrix elements of the difference of operators get an oscillatory behavior around a finite value, smaller than the one obtained for the corresponding diagonal matrix element. The amplitude of these oscillations is generically small and depends on the value of $q_{\varepsilon}$ and the considered anisotropy profile, being negligible when} so are $q_{\varepsilon}^{n-1}\partial^{n}f(\Lambda')/n!$ (for all $n\ge2$) compared with $\partial f(\Lambda')$.

The discussed approximation for the anisotropy term, in particular, allows us to obtain a manageable form for the Hamiltonian constraint of the Bianchi I model, which can now be rewritten in the approximate form:
\begin{align}\label{constBI}
\hat{C}_{\text{BI}}\simeq\hat{C}_{\text{FRW}}-\frac{\pi G\hbar^2}{4}\hat{\tilde\Omega}\hat{\Theta}'.
\end{align}
This must be a good approximation for states in the ``high-energy'' sector of the FRW-geometry---where the only relevant contributions correspond to $\rho\gg 10$---and whose anisotropy profile is smooth enough so as that the dominant contribution to its variation up to the scale $q_{\varepsilon}$ is proportional to its increment at that scale. Notice that, in each eigenspace of the operator $\hat{\Theta}'$, the eigenvalue equation of the approximate constraint obtained for Bianchi I reduces to a difference equation in one dimension, coordinatized by $v\in \mathcal{L}^+_\varepsilon$. In this sense, solving that difference equation amounts to solving the constraint for the Bianchi I model in our approximation.

\subsection{Approximation 3: Dealing with the interaction and the approximate anisotropy terms}

The approximate anisotropy term $2\hat{\tilde\Omega}\hat{\Theta}'$ introduces a new complication in the resolution of our constraint, because $\hat{\tilde\Omega}$ does not have a diagonal action on the eigenstates of $\hat{\Omega}^2$, and $\hat{\Theta}'$ does not commute with $\widehat{e^{\pm2\Lambda}}$. This is no fundamental obstruction in the absence of inhomogeneities, as we have commented above. However, when non-zero modes are present, we cannot reduce the eigenvalue problem associated with the constraint to a manageable one-dimensional problem. Nonetheless, let us recall that we are interested in regimes with small anisotropy effects. For this, we focus our discussion on physical states for which the expectation value of $\hat{\Theta}' $ is small, and can be disregarded. On the other hand, we can adopt a similar strategy to deal with the interaction term of the constraint. Since it comes multiplied by $\widehat{e^{-2\Lambda}}$, we can choose our profiles in $\Lambda$ in such a way that the expectation value of $\widehat{e^{-2\Lambda}}$ is also negligible, and can be ignored in comparison with the rest of contributions that arise in the constraint equation. Besides we assume that, for our states, non-diagonal terms can be disregarded in $\hat{\Theta}' $ and $\widehat{e^{-2\Lambda}}$. This must be the case for sufficiently peaked Gaussian profiles (in $\Lambda$ and the anisotropy momentum represented by $\hat{\Theta}'$), on which we will concentrate our attention from now on.

It is worth emphasizing that the reason why the anisotropy term in the Hamiltonian constraint is small for this class of states is because they are peaked around a vanishing expectation value of $\hat{\Theta}' $, that is, they are peaked on the sector of small values of the momentum of the anisotropy variable  $\Lambda$. However, this does not imply by any means that those states are peaked around trajectories describing isotropic geometries. On the contrary, isotropic trajectories would be classically characterized by the relation $3\Lambda= \log(v/2)$, as we explained in Subsec. \ref{sec:HGM1}. The states with Gaussian profiles that we are considering are rather peaked around a constant value of $\Lambda$, totally independent of the value of $v$. Even so, they provide an anisotropy contribution to the constraint that is negligible. Moreover, as we have commented, we focus on the case where the constant peak value of $\Lambda$ is very large, so as to be able to disregard as well the term that goes with $\widehat{e^{-2\Lambda}}$.

To seek for states with the desired properties, let us first carry out the spectral analysis of the operator $\hat{\Theta}' $. One can easily prove that $\hat{\Theta}' $ is (essentially) self-adjoint. Its spectrum is absolutely continuous, bounded, doubly-degenerate, and equal to the interval $[-4/q_\varepsilon,4/q_\varepsilon]$. Given an eigenvalue $s$, the corresponding eigenfunctions are
\begin{align}
e_s^{(1)}(\Lambda)\equiv \langle \Lambda|e_s^{(1)}\rangle=N(s) e^{i\frac{\Lambda }{q_\varepsilon}x(s)}, \quad e_s^{(2)}(\Lambda)\equiv \langle \Lambda|e_s^{(2)}\rangle=N(s) e^{i\frac{\Lambda }{q_\varepsilon}[\pi-x(s)]},
\end{align}
where
\begin{align}
x(s)=\arcsin\left(s \frac{q_\varepsilon}{4}\right),\qquad N(s)=\frac{1}{\sqrt{2\pi}}\sqrt[4]{\frac{q_{\varepsilon}^2}{16-s^{2}q_{\varepsilon}^2}}.
\end{align}
These eigenfunctions provide a resolution of the identity in $\mathcal{H}^{q_\varepsilon}_{\Lambda^\star}$, given by
\begin{align}
\mathbb{I}_ {\mathcal{H}^{q_\varepsilon}_{\Lambda^\star}}=  \int_{-4/q_{\varepsilon}}^{4/q_{\varepsilon}} ds
\left(|e_s^{(1)}\rangle \langle e_s^{(1)}|+ |e_s^{(2)}\rangle \langle e_s^{(2)}|\right).
\end{align}

In the rest of our discussion, we will consider states whose anisotropy part is given by the Gaussian-like profile:
\begin{align}\label{ani}
|{\psi_{\bar{\Lambda}}}\rangle&= \int_{-4/q_{\varepsilon}}^{4/q_{\varepsilon}} ds \sqrt{\frac{q_\varepsilon}{4\sigma_{s}\sqrt{\pi}\cos[x(s)]}}e^{-\frac{x^2(s)}{2\sigma^{2}_{s}}-ix(s)\frac{\bar\Lambda}{q_\varepsilon}}  |e_s^{(1)}\rangle\nonumber\\
&\simeq \sum_{\Lambda\in\mathcal{L}^{q_\varepsilon}_{\Lambda^\star}} \frac{\sqrt{\sigma_{s}}}{\sqrt[4]{\pi}} e^{-\frac{\sigma^{2}_{s}}{2q_{\epsilon}^{2}}(\Lambda-\bar{\Lambda})^{2}}|\Lambda\rangle.
\end{align}
The last approximation is valid as long as $\sigma_{s}\ll\pi/2$. These states are peaked around the anisotropy value $\bar{\Lambda}$ and around the momentum $s=0$. A straightforward calculation shows indeed that, on the one hand, $\langle {\psi_{\bar{\Lambda}}}| \hat{\Theta}' |{\psi_{\bar{\Lambda}}}\rangle=0$, as expected for Gaussian-like profiles peaked around a vanishing anisotropy momentum. On the other hand,
\begin{align}
\langle{\psi_{\bar{\Lambda}}}|\widehat{e^{-2n\Lambda}}|{\psi_{\bar{\Lambda}}}\rangle & \simeq e^{-2n\bar{\Lambda}+n^{2}\frac{q_{\epsilon}^{2}}{\sigma^{2}_{s}}},
\end{align}
where $n$ can be any integer. Therefore,  if $\bar\Lambda\gg q_{\varepsilon}^2/\sigma_s^2$, we have $\langle{\psi_{\bar{\Lambda}}}|\widehat{e^{-2\Lambda}}|{\psi_{\bar{\Lambda}}}\rangle\ll 1$, and similarly for the dispersion:
\begin{align}
\sqrt{| \langle{\psi_{\bar{\Lambda}}}|\widehat{e^{-4\Lambda}}-( \langle{\psi_{\bar{\Lambda}}}|\widehat{e^{-2\Lambda}}|{\psi_{\bar{\Lambda}}}\rangle)^2|{\psi_{\bar{\Lambda}}}\rangle|}=\sqrt{e^{-4\bar{\Lambda}+4\frac{q_{\epsilon}^{2}}{\sigma^{2}_{s}}}\left(1-e^{-2\frac{q_{\epsilon}^{2}}{\sigma^{2}_{s}}}\right)} \ll 1.
\end{align}

In conclusion, by considering states of the form given above, with a small value of the Gaussian width $\sigma_s$ and a sufficiently large value of the anisotropy peak $\bar\Lambda$, the expectation values (with respect to the anisotropy dependence) of both the anisotropy term and the interaction term in the Hamiltonian constraint are negligible (assuming a reasonable content of inhomogeneities). Hence, we can approximate the constraint by the solvable one
\begin{align}\label{const}
\hat{C}_{\text{app}}=\hat{C}_{\text{FRW}}+\frac{2\pi G \hbar^{2}}{\beta} \widehat{e^{2\Lambda}}\hat{H}_0=-\frac{3\pi G\hbar^2}{8}\hat{\Omega}^2-\frac{\hbar^2\partial_\phi^2}{2}+\frac{2\pi G \hbar^{2}}{\beta}\widehat{e^{2\Lambda}}\hat{H}_0.
\end{align}

\section{Approximate solutions to the Gowdy model}
\label{sec:sol}

Let us now solve the constraint $(\Psi|\hat{C}_{\text{app}}^{\dagger}=0$. We consider generalized states
\begin{equation}
(\Psi|=\int_{-\infty}^{\infty} d{\phi}\sum_{v\in\mathcal{L}^+_{\varepsilon}}\sum_{\Lambda\in\mathcal{L}^{q_\varepsilon}_{\Lambda^\star}}
\sum_{\mathfrak{n}^{\xi},\mathfrak{n}^{\varphi}}\Psi(\phi,v,\Lambda,\mathfrak{n}^{\xi},\mathfrak{n}^{\varphi})
\langle{\phi},v,\Lambda,\mathfrak{n}^{\xi},\mathfrak{n}^{\phi}|,
\end{equation}
where
\begin{align}
\langle{\phi},v,\Lambda,\mathfrak{n}^{\xi},\mathfrak{n}^{\phi}|=\langle \phi|\otimes \langle v| \otimes \langle\Lambda| \otimes \langle \mathfrak{n}^{\xi}, \mathfrak{n}^{\phi}|,
\end{align}
the bra denoting a state in the corresponding dual space.
We can write
\begin{align} \label{profil}
\Psi(\phi,v,\Lambda,\mathfrak{n}^{\xi},\mathfrak{n}^{\varphi})=\int_{-\infty}^{\infty}\!\! dp_{\phi} \int_{0}^{\infty}\!\!\! d\rho\,  \Psi(p_\phi,\rho,\Lambda,\mathfrak{n}^{\xi},\mathfrak{n}^{\varphi}) e^{\varepsilon}_\rho(v) e_{p_\phi}(\phi).
\end{align}
On these states the constraint $(\Psi |\hat{C}_{\text{app}}^{\dagger}|\phi,v,\Lambda,\mathfrak{n}^{\xi},\mathfrak{n}^{\phi}\rangle=0$ reads
\begin{align}
\int_{-\infty}^{\infty} \!\!dp_{\phi}\int_{0}^{\infty} \!\!\!d\rho\,  \Psi(p_\phi,\rho,\Lambda,\mathfrak{n}^{\xi},\mathfrak{n}^{\varphi}) e^{\varepsilon}_\rho(v) e_{p_\phi}(\phi)\left(-\frac{3\pi G\hbar^2}{8}\rho^{2}+\frac{p_{\phi}^{2}}{2}+\frac{2\pi G \hbar^{2}}{\beta} e^{2\Lambda}H_{0}(\mathfrak{n}^{\xi},\mathfrak{n}^{\varphi})\right)=0,
\end{align}
where
\begin{align}
H_{0}(\mathfrak{n}^{\xi},\mathfrak{n}^{\varphi})\equiv\langle \mathfrak{n}^{\xi},\mathfrak{n}^{\varphi}| \hat{H}_0|\mathfrak{n}^{\xi},\mathfrak{n}^{\varphi}\rangle=\sum_{m\in\mathbb{Z}-\{0\}}|m|(n^\xi_m+n^\varphi_m).
\end{align}
Note that $H_0(\mathfrak{n}^{\xi},\mathfrak{n}^{\varphi})$ is always non-negative.
Solving for $\rho$, the solutions of the constraint are states with profiles \eqref{profil} of the form
\begin{align}
\Psi(\phi,v,\Lambda,\mathfrak{n}^{\xi},\mathfrak{n}^{\varphi})= \int_{-\infty}^{\infty} dp_{\phi}\, \Psi(p_\phi,\Lambda,\mathfrak{n}^{\xi},\mathfrak{n}^{\varphi}) e^{\varepsilon}_{\rho(p_\phi,\Lambda,\mathfrak{n}^{\xi},\mathfrak{n}^{\varphi})}(v) e_{p_\phi}(\phi),
\end{align}
where
\begin{align}\label{rhosol}
\rho(p_\phi,\Lambda,\mathfrak{n}^{\xi},\mathfrak{n}^{\varphi})=\sqrt{\frac{4}{3\pi G\hbar^2}p_\phi^{2}+\frac{16}{3\beta} e^{2\Lambda}H_{0}(\mathfrak{n}^{\xi},\mathfrak{n}^{\varphi})}.
\end{align}
Let us emphasize that the term inside the square root is non-negative, so that the solution for $\rho$ is well defined.

It is now straightforward to construct approximate solutions to the Hamiltonian constraint of the Gowdy model. In order to do that, let us first restrict the solutions $(\Psi|$ so that their anisotropy dependence is given by the Gaussian wavefunctions studied in the previous section:
\begin{align}
\Psi(p_\phi,\Lambda,\mathfrak{n}^{\xi},\mathfrak{n}^{\varphi})=\Psi(p_\phi,\mathfrak{n}^{\xi},\mathfrak{n}^{\varphi})\psi(\Lambda),
\end{align}
with [see Eq. \eqref{ani}]
\begin{align}\label{prof}
\psi(\Lambda)=\frac{\sqrt{\sigma_{s}}}{\sqrt[4]{\pi}} e^{-\frac{\sigma^{2}_{s}}{2q_{\epsilon}^{2}}(\Lambda-\bar{\Lambda})^{2}},\quad \sigma^{2}_{s}\ll \frac{\pi}{2},\quad  \bar\Lambda\gg \frac{q_{\varepsilon}^2}{\sigma^{2}_{s}}.
\end{align}
We also assume a sufficiently small content of inhomogeneities, in the sense that their interaction term can be ignored for the kind of anisotropy profiles under consideration.
Besides, since we want that all the introduced approximations are well justified, we need to restrict to considerably large values of $\rho$ in Eq. \eqref{rhosol}, $\rho\gg 10$. Taking into account the expression of $\rho(p_\phi,\Lambda,\mathfrak{n}^{\xi},\mathfrak{n}^{\varphi})$ and recalling that the contribution of $H_0$ is non-negative, we can guarantee that $\rho\gg 10$ by restricting to scalar field momenta such that $p_{\phi}^2\gg 75 \pi G\hbar^2 \approx 200 G\hbar^2$.

Once we have dealt with the Hamiltonian constraint satisfactorily, we still have to impose the diffeomorphisms constraint \eqref{dif}. This last constraint restricts the numbers of particles so that
\begin{align}
\sum_{m\in\mathbb{N}^{+}}m\left(n^\xi_m+n^\varphi_m-n^\xi_{-m}-n^\varphi_{-m}\right)=0.
\end{align}
States in the sector $p_{\phi}^2\gg 200 G\hbar^2$ with the anisotropy profiles \eqref{prof} and a particle content that satisfies the above condition must be approximate solutions for the Gowdy model (provided that the particle interaction is not overwhelmingly large).

We conclude by suggesting a convenient choice for the parameter $q_\varepsilon$. In principle, there is a big freedom in fixing this parameter. Nonetheless, instead of doing an arbitrary choice, we are going to motivate the selection of a specific value for it employing physical arguments.
The motivation comes from the fact that the homogeneous scalar field momentum is a constant of motion and provides a natural scale in the system (in each of its generalized eigenspaces, where the approximation to the Hamiltonian constraint can be carried out independently). We have already discussed that, if the gravitational FRW contribution is only relevant in a sector $\rho\geq \rho^{\star}$, the optimal choice of $q_\varepsilon$ would be the smallest element in $\mathcal{W}_\varepsilon$ larger than $\log{(1+4/\rho^{\star})}$. As we have just seen, the momentum $p_{\phi}$ provides in fact a lower bound on $\rho$, via Eq. \eqref{rhosol}. Hence, it seems natural to make the choice $q_\varepsilon=\log\left(1+2/v^\star\right)$ where $v^{\star}$ is the point in the semilattice $\mathcal{L}^+_\varepsilon$ smaller but closest to the lower bound on $\rho/2$, namely,
\begin{align}
v^\star={\rm max}\left\{v\in\mathcal{L}^+_\varepsilon \text{ such that } v<  \frac{|p_\phi|}{\sqrt{3\pi G\hbar^2}} \right\}.
\end{align}
Note that, for large values of the homogeneous scalar field momentum, the parameter $q_\varepsilon$ selected in this way is really small.

\section{Conclusions}
\label{sec:C}

We have studied approximation methods in the context of LQC in order to construct physical solutions of inhomogeneous and anisotropic systems. We have applied those methods to the hybrid quantization of the $T^{3}$ Gowdy model with linear polarization and a minimally coupled massless scalar field, with the same symmetries as the spacetime metric. This model can be regarded as an FRW background with anisotropies and (matter and gravitational) inhomogeneities propagating on it, inasmuch as the Hamiltonian constraint of the system is given by the Hamiltonian constraint of the FRW model coupled to a homogeneous massless scalar, an anisotropy term (which, together with the FRW contribution, gives the Hamiltonian constraint of the Bianchi I model), and two inhomogeneous terms (one gives the free energy of the inhomogeneities and the other is an interaction term). The last three terms are all coupled to the (homogeneous and isotropic) FRW background.

One of the main complications when trying to solve the Gowdy Hamiltonian constraint is the presence of the anisotropy term. On the one hand, it does not commute with the FRW Hamiltonian constraint. On the other hand, its action on the anisotropy variable is quite complicated, in practice making unviable its spectral analysis. We have shown how to approximate this operator by a tensor product of two operators, one that only acts in the volume variable $v$ and another operator that only acts in the anisotropy label $\Lambda$. The latter is a simple discrete derivative which retains the discrete effects coming from the loop quantization of the anisotropies, but with a simpler domain of definition and action than those of the original anisotropy operator. As a consequence, the original anisotropy superselection sector splits in subspaces corresponding to states with support on equi-spaced lattices. The spectrum of the new discrete derivative operator in each of these subspaces can be characterized completely. With this characterization, and using our approximation for the anisotropy term, we obtain a Hamiltonian constraint for the Bianchi I model whose resolution reduces to a one-dimensional problem. In turn, we are also able to find states whose profiles in $\Lambda$ are sufficiently peaked, both in the anisotropy and in its momentum---given by the introduced discrete derivative---, as to allow us to disregard in the Hamiltonian constraint not only this anisotropy term but also the contribution of the interaction term of the inhomogeneities. By considering such profiles in $\Lambda$, we obtain approximate solutions to the Gowdy model corresponding to states with a small quantum effect of the anisotropies and a reasonable content of inhomogeneities. The larger the momentum of the homogeneous massless scalar (which is a constant of motion), the better these solutions must approximate those of the Gowdy model. We leave for future research the analysis of the time evolution of the approximate solutions obtained in this work.

This is not the first time that a model which behaves quantum mechanically as FRW plus inhomogeneities is analyzed in the context of LQC. Actually, the hybrid quantization employed here has also been applied in the  quantization of FRW models with cosmological perturbations \cite{inf-hybrid,inf-hybrid2,inf-ash}. We hope that the approximation methods developed in this work can serve as well to find approximate solutions in even more realistic models, like those describing inflationary scenarios.

\section{Acknowledgements}
The authors are grateful to E. Livine, J. Olmedo and P. Tarr\'{\i}o for discussions. This work was partially supported by the Spanish MICINN/MINECO Project No. FIS2011-30145-C03-02. D M-dB is supported by CSIC and the European Social Fund under the Grant No. JAEPre-09-01796. He also wants to thank Perimeter Institute for warm hospitality during the preparation of this work. Research at Perimeter Institute is supported by the Government of Canada through Industry Canada and by the Province of Ontario through the Ministry of Research and Innovation.

\end{document}